\pdfoutput=1
\documentclass[twocolumn,showpacs,preprintnumbers,amsmath,amssymb]{revtex4}

\usepackage{xspace}
\usepackage{epsfig}
\usepackage{amsfonts}
\usepackage{graphicx}
\usepackage{url}
\usepackage[dvipdfm,%
              colorlinks,%
          linkcolor=blue,%
        citecolor = blue,%
              hyperindex,%
        plainpages=false,%
         urlcolor = blue,%
       pdfstartview=FitH]{hyperref}

\begin{document}


\title{\textsf{Effect of optical phonon scattering on the performance of GaN transistors}}

\author{Debdeep Jena}
\affiliation{Electrical Engineering, University of Notre Dame, IN, 46556, USA}
\author{Siddharth Rajan}
\affiliation{Electrical and Computer Engineering, The Ohio State University, Columbus, OH, 43210, USA}

\date{\today}

\begin{abstract}
A model based on optical phonon scattering is developed to explain peculiarities in the current drive, transconductance, and high speed behavior of short gate length GaN transistors.  The model is able to resolve these peculiarities, and provides a simple way to explain transistor behavior in any semiconductor material system in which electron-optical phonon scattering is strong.
\end{abstract}



\maketitle


GaN-based high electron mobility transistors (HEMTs) exhibit a number of peculiar features.  Polarization induces some of the highest channel 2D electron gas (2DEG) sheet charge densities in all semiconductors, with values in the range of $n_{s} \sim 1- 4 \times 10^{13}$/cm$^{2}$.  If an effective electron saturation velocity of $v_{sat} \sim 10^{7}$cm/s is assumed, the drain current per unit device width ($W$) is expected to be $I_{d}^{sat}/W = J_{sat} = q n_{s} v_{sat} \sim 5$ A/mm for $n_{s} \sim 3 \times 10^{13}$/cm$^{2}$.  However, experimentally measured saturation currents fall far short of such values - which is the first peculiar feature.  Second, the `gain' of GaN HEMTs, or their transconductance given by $g_{m} = \partial{I_{d}}/\partial{V_{gs}}$ shows a sharp drop after reaching a maximum as the gate voltage is increased.  This is also not expected, since the constant saturation velocity model predicts $g_{m}/W = C_{gs} v_{sat}$, which should stay constant at high gate voltages.  Third, the current-gain cutoff frequency ($f_{T}$) also follows the trend of $g_{m}$.  Earlier explanations attribute this to a variable source resistance \cite{PalaciosRa}.  However, recent results \cite{drc10hrl} with negligible source access resistances ($\sim 50$ $\Omega \cdot \mu$m) show an even pronounced `peaky' behavior of $g_{m}$ and $f_{T}$ vs $V_{gs}$.  The purpose of this work is to offer a novel, yet simple model of GaN transistor operation that reconciles all three peculiarities.  In addition, the model also brings forth a remarkably simple picture of transistor operation when electron-optical phonon scattering is strong, one that can be used for quantitative predictions.  It also highlights the importance of an accurate representation of the current drive in a transistor, a central quantity that determines all other device performance characteristics.

The first hint at resolving the above peculiarities appears upon a close examination of the strength of electron-optical phonon interactions in GaN.  Owing to the light mass of the nitrogen atom (lightest in all III-V semiconductors), the polar optical phonon energy is high.  Combined with the high electronegativity of nitrogen, it results in a very high scattering rate.  The mean free path of energetic electrons emitting optical phonons is $\lambda_{op} \sim a_{B}^{\star} \epsilon_{\infty}/(\epsilon_{0} - \epsilon_{\infty}) \sim 3.5$ nm ($a_{B}^{\star}$ is the effective Bohr-radius and $\epsilon_{0}, \epsilon_{\infty}$ are the static and high-frequency dielectric constants).  This is much shorter than most other semiconductors (Si, GaAs). It prevents ballistic transport at high-fields, and damps velocity overshoot effects even for very short gate length GaN devices.  Similar hi-field phenomena also occurs for other high phonon energy materials such as carbon nanotubes and graphene \cite{prl00dekkerCNTsatCurrent, jap09djCNTsatCurrent, drc08grJsat}.

\begin{figure}
\begin{center}
\leavevmode \epsfxsize=3.2in \epsfbox{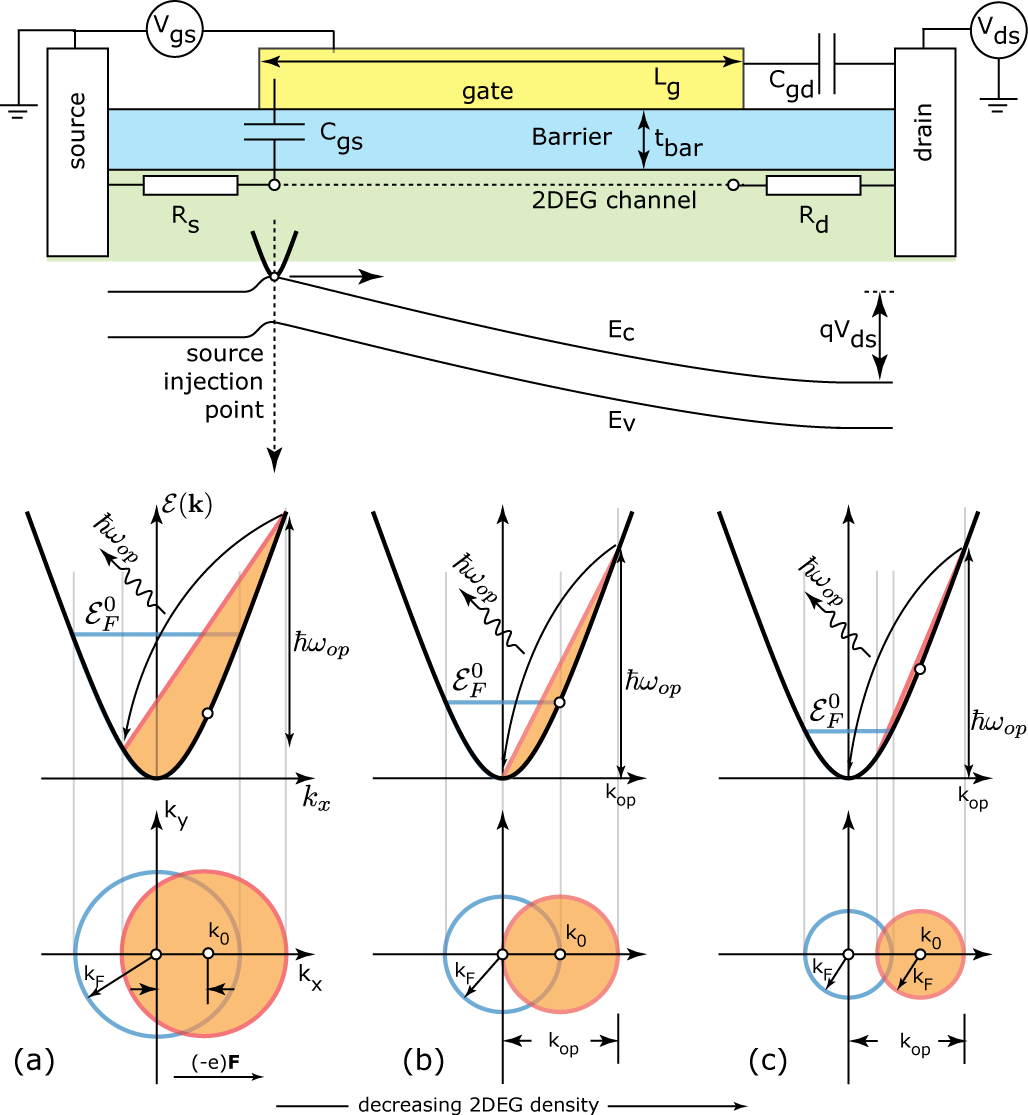}
\caption{Transistor cross section, energy band diagram, bandstructure, and ${\bf k}-$space occupation of electrons.} 
\label{Fig1}
\vspace{-7ex}
\end{center}
\end{figure}

Fig \ref{Fig1} depicts the band diagram of a typical GaN transistor.  To understand the characteristics of a short gate-length device, we inspect the electron distribution at the `source injection point' \cite{jap94natori}.  The energy and the ${\bf k}-$space distribution of 2DEG electrons at this point is depicted for various carrier densities dictated by the gate voltage and the gate-to-source injection plane capacitance.  We assume an electrostatically well-designed device with negligible drain-induced barrier lowering and other short channel effects.  Fig \ref{Fig1}(a, b, and c) depict the carrier distributions at equilibrium (Fermi circles of radius $k_{F}=\sqrt{2 \pi n_{s}}$ at $V_{ds}=0$, $n_{s}$ is the 2d sheet density), and at current saturation (shifted Fermi circle of the same radius).  In a ballistic transistor \cite{jap94natori}, the difference of quasi-Fermi levels of the right- and the left-going carriers is the drain bias.  However, in GaN transistors, ultrafast optical phonon emission locks this difference at the optical phonon energy ($\hbar \omega_{op} \sim 92$ meV), as shown in Fig \ref{Fig1}.  The carrier distribution in ${\bf k}-$space at current saturation is depicted by the filled circles at various 2DEG carrier concentrations; as the gate pinches the channel off at the source injection point, the Fermi circle shrinks.  

At high 2DEG densities (Fig \ref{Fig1}(a)), electrons from the highest right-going energy state emit an optical phonon and scatter into the highest empty left-going state.  Scattering into the bottom of the subband is Pauli-blocked due to the high degeneracy, but becomes possible when the condition $2k_{F} \leq k_{op}$ is met as shown in Fig \ref{Fig1}(b).   From Fig \ref{Fig1}(b), we get $\hbar^{2}k_{op}^{2}/2 m^{\star} = \hbar \omega_{op}$, and the crossover 2DEG density is $n_{0} = k_{op}^{2}/8 \pi \sim 1.9 \times 10^{12}$/cm$^{2}$ (here $m^{\star} \sim 0.2 m_{0}$ is the electron effective mass of GaN).  For $n_{s} \leq n_{0}$, the highest energy right-going state has an energy $\hbar \omega_{op}$, and the lowest occupied state also moves to the right, as seen in Fig \ref{Fig1}(c).  Now that the carrier distribution at the source injection point in the ${\bf k}-$space is identified, the current flowing in the HEMT can be calculated.
The current is found by summing over the group velocities of all occupied states in the ${\bf k}-$space; for 2D carriers in a parabolic band, the result is simple:
\begin{equation}
{\bf J} = \frac{2q}{(2\pi)^{2}} \int_{\Gamma} d^{2}k (\frac{\hbar {\bf k}}{m^{\star}}) f({\bf k}) = q n_{s} (\frac{\hbar {\bf k}_{0}}{m^{\star}}),
\end{equation}
where ${\bf k}_{0}$ is the geometrical {\em centroid} of the occupied area in the ${\bf k}-$space.  The result is {\em general}, valid at all temperatures if $f({\bf k})$ is symmetric around ${\bf k}_{0}$, and is not restricted to GaN transistors.  For the specific case considered here, ${\bf k}_{0} = k_{0} \hat{x}$, and the scalar drain current is $J = q n_{s} (\hbar k_{0} / m^{\star})$.  Thus, the net current per unit width under any bias conditions can be found if the centroid of the distribution in ${\bf k}-$space is identified.

\begin{figure}
\begin{center}
\leavevmode \epsfxsize=3.0in \epsfbox{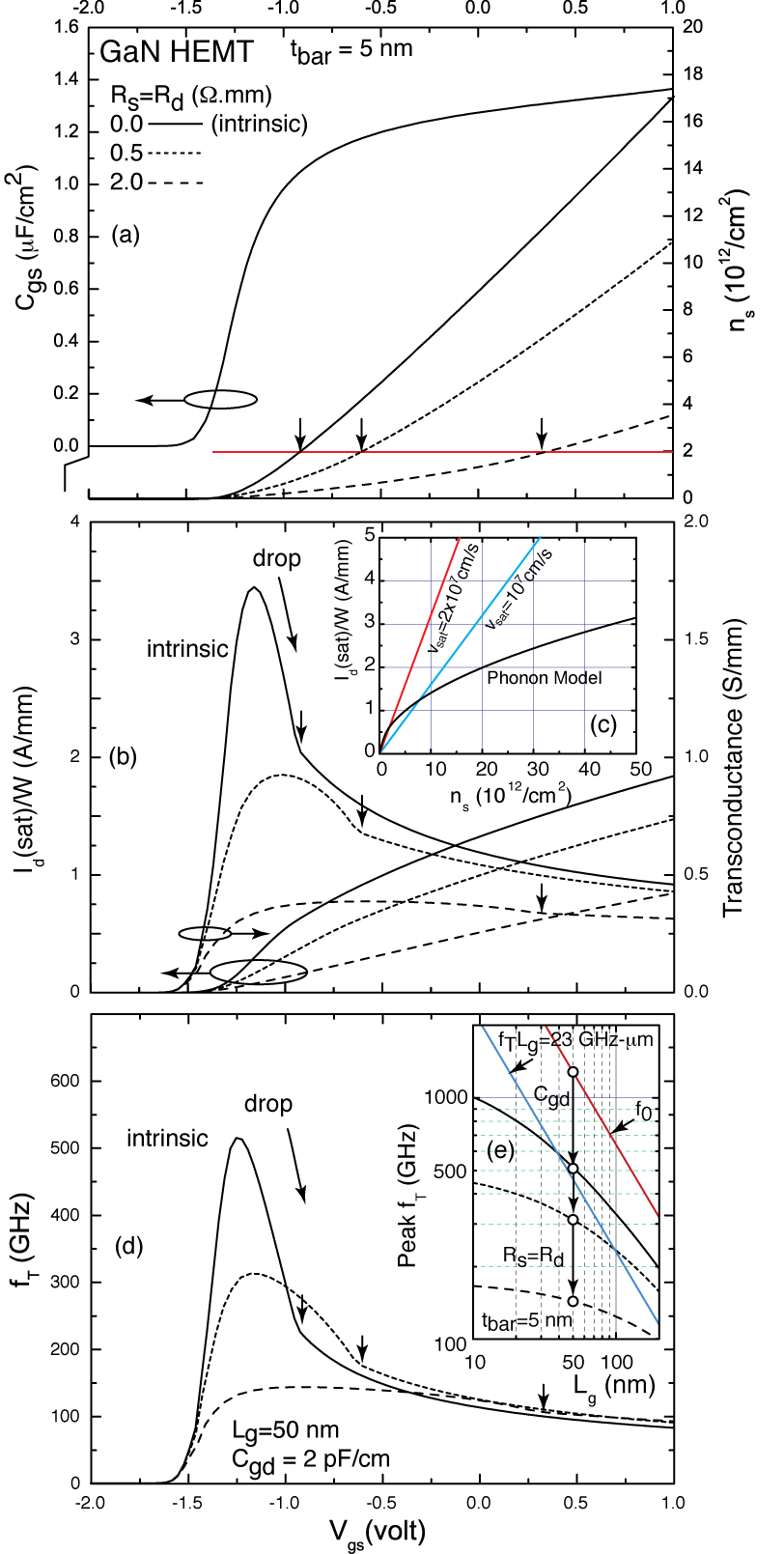}
\caption{AlGaN/GaN transistor characteristics calculated using the phonon model.} 
\label{Fig2}
\vspace{-7ex}
\end{center}
\end{figure}

For high sheet densities ($n_{s} > n_{0} \sim 1.9 \times 10^{12}$/cm$^{2}$ as depicted in Fig \ref{Fig1}(a)), the centroid is found by solving
\begin{equation}
\frac{\hbar^{2} (k_{F} + k_{0})^{2}}{2 m^{\star}} - \frac{\hbar^{2} (k_{F} - k_{0})^{2}}{2 m^{\star}} = \hbar \omega_{op},
\end{equation}
yielding $k_{0} = m^{\star} \omega_{op}/2 \hbar k_{F}$.  Similarly, for the low-carrier density condition in Fig \ref{Fig1}(c), it is easily seen that the centroid is at $k_{0} = k_{op} - k_{F}$.  The HEMT saturation current density is thus given by 
\begin{eqnarray}
J  = 	&	q n_{s} \frac{\hbar (k_{op} - \sqrt{2 \pi n_{s}})}{m^{\star}}	& n_{s} \leq n_{0}, \label{jsatL} \\
& = \frac{q}{\sqrt{8 \pi}} \omega_{op} \sqrt{n_{s}} 			& n_{s} > n_{0}. \label{jsatH}
\end{eqnarray}
The results indicate a significant deviation from a constant saturation velocity model.  The current is proportional to the {\em square root} of the 2DEG density in the high density (and high current) regime.  The difference is highlighted in Fig \ref{Fig2}(c).

The transconductance in S/m units is given by $g_{m} = \partial J / \partial V_{g} = (\partial J / \partial n_{s}) C_{gs} $, where $C_{gs}$ is the intrinsic capacitance per unit area between the gate metal and the source injection point.  Using Eqs. (\ref{jsatL}, \ref{jsatH}), we get
\begin{eqnarray}
g_{m} = & \frac{\hbar}{m^{\star}}(k_{op} - \frac{3}{2}\sqrt{2 \pi n_{s}}) C_{gs}	& n_{s} \leq n_{0} \label{gmL} \\
& = \frac{\omega_{op}}{4\sqrt{2 \pi n_{s}}} C_{gs} 	& n_{s} > n_{0} \label{gmH},
\end{eqnarray}
which casts the transconductance in a familiar form as the product of the gate capacitance and the prefactor indicative of an effective electron ensemble velocity.  Note that the effective velocity is strongly dependent on the carrier density, and therefore we are forced to refrain from invoking the concept of a saturation velocity that is a material constant.  Furthermore, it is evident that the transconductance at high carrier densities will {\em decrease} as $g_{m} \sim 1/\sqrt{n_{s}}$.  With a parasitic source resistance $R_{s}$ depicted in Fig \ref{Fig1}, there is a voltage drop from the source contact to the source injection point, and the measured transconductance becomes $g_{m}^{ext} = g_{m}/(1 + R_{s}g_{m})$.  

The speed of the transistor can be evaluated by either linearizing the small-signal temporal drain current (Eqs. \ref{jsatL}, \ref{jsatH}) response to a sinusoidal gate voltage sweep, or by a straightforward charge control analysis; both methods lead to the same intrinsic channel delay.  The delay is $\tau_{c} = \partial Q / \partial J = q L_{g} \partial n_{s}/\partial J$, which yields
\begin{eqnarray}
\tau_{c} = & \frac{  m^* }{\hbar\left(k_{\text{op}}-\frac{3}{2}\sqrt{2 \pi  n_s}\right) }L_g  \hspace{20pt} & n_s \leq n_0 \label{tauL} \\ 
 &   = \frac{ 4 \sqrt{2 \pi  n_s}}{\omega _{\text{OP}}}L_g          & n_s > n_0 \label{tauH}.
\end{eqnarray}
We note that the intrinsic channel delay is of the form $\tau_{c} = L_{g}/v_{eff}$, where the effective velocity is strongly dependent on the 2DEG carrier concentration controlled by the gate voltage.  The unity current-gain cutoff frequency of the transistor is then calculated as $f_{T} = 1/2 \pi \tau_{tot}$, where $\tau_{tot} = \tau_{c} + C_{gd}/g_{m} + C_{gd}(R_{s} + R_{d})$, where $C_{gd}$ is the gate-drain overlap capacitance, and $R_{s}, R_{d}$ are the source and drain resistances (units- $\Omega.$mm).  All these lumped elements are depicted in Fig \ref{Fig1}.  The transconductance is the `intrinsic' $g_{m}$ from Eqs. \ref{gmL}, \ref{gmH}.  The dependence on $g_{m}$ indicates that in the high density regime, $f_{T}$ will decrease with increasing channel charge or equivalently, the gate voltage.  

The transistor characteristics under current saturation are now captured by Eqs. \ref{jsatL} - \ref{tauH}.  The only parameter necessary to quantitatively investigate particular cases is the gate-channel capacitance $C_{gs}$.  To model a realistic GaN HEMT, we use a self-consistent Poisson-Schrodinger solver to calculate $C_{gs}$ as a function of $V_{gs}$.  The calculation gives an accurate dependence of charge at the source injection point on the gate voltage.  The calculation accounts for quantization effects in the channel 2DEG.  The device structure chosen for highlighting the phonon saturation model is an Al$_{0.35}$Ga$_{0.65}$N/GaN HEMT with a $t_{bar}= 5 $ nm barrier.  Fig \ref{Fig2} shows the device characteristics using Eqs. \ref{jsatL} - \ref{tauH} using the self-consistent Poisson-Schrodinger results as the input.
The calculated $C_{gs}$ is shown as a function of the gate voltage in Fig \ref{Fig2}(a), indicating a threshold voltage of $\sim -1.5$ V.  This figure also shows the 2DEG density at the source injection point for three different $R_{s}$ values.  This sheet density depends on the potential difference between the gate and the source-injection point ($V_{gs}^{0}$).  For $R_{s} \neq 0$, the external gate voltage is $V_{gs} = V_{gs}^{0} + JR_{s}$, and this is what is plotted in Fig \ref{Fig2}.  As the source-resistance increases, a higher gate voltage is necessary to induce the same 2DEG charge.  The specific case for $n_{s} = n_{0}$ is highlighted by a red line and three arrows in Fig \ref{Fig2}(a).  This is also the `crossover' density at the source injection point as identified earlier.  The 2DEG density at the source injection point is $n_{s} \sim 10^{13}$/cm$^{2}$ at $V_{gs}=1.0$ V if $R_{s}=0.5$ $\Omega.$mm.

The drain current and the transconductance are shown in Fig \ref{Fig2}(b) as a function of the gate bias.  The intrinsic transconductance peaks at $g_{m} \sim 1.7 $ S/mm.  In addition, there is an inflection point at the gate voltage which corresponds to $n_{s} = n_{0}$ at the source injection point.  The inflection points are indicated by arrows.  Experimental $g_{m}$ values as high as $0.8 - 0.9$ S/mm have been reported recently.  The phonon model thus predicts that as $R_{s}$ in GaN HEMTs is lowered, the $g_{m}$ will become more `peaky'.  The inset (Fig \ref{Fig2}(c)) shows saturated drain current as calculated from the phonon model (Eqs. \ref{jsatH}, \ref{jsatL}) in comparison with constant saturation velocity models.  The phonon model follows a $\sqrt{n_{s}}$ dependence at high densities (for example $\sim 2$ A/mm at $n_{s} \sim 2 \times 10^{13}$/cm$^{2}$), and therefore explains experimentally observed currents in GaN transistors.

The calculated current gain cutoff frequencies are shown in Fig \ref{Fig2}(d) for $L_{g} = 50$ nm using $C_{gd} = 2$ pF/cm.  The gate dependence of $f_{T}$ follows that of $g_{m}$, replete with a maximum near threshold voltage, inflection points, and sharp drop at high gate voltages.  Severe degradation of $f_{T}$ and `smoothening' of the sharp features occurs with increasing $R_{s}$ and $R_{d}$.  To study the gate length dependence of $f_{T}$, we choose the peak $f_{T}$ for each gate length and plot it against $L_{g}$ in the inset Fig \ref{Fig2}(e) for $t_{bar} = 5$ nm for the intrinsic case ($f_{0} = 1/2 \pi \tau_{c}$ as $n_{s} \rightarrow 0$), with $C_{gd}$, and with access resistances.  A $f_{T} \cdot L_{g} = 23 $ GHz.$\mu$m line corresponding to a constant electron saturation velocity of $v_{sat} \sim 1.4 \times 10^{7}$ cm/s is plotted for comparison \cite{jessen07}.  The phonon model predicts a slower increase in $f_{T}$ with scaling than the velocity saturation model for small gate lengths.  

The model presented here is able to reconcile the three peculiarities of GaN transistors and explains the current drive, $g_{m}$, and $f_{T}$ behavior with gate bias.  The limitation of the model as presented is that it assumes single subband occupation, which can be violated for high 2DEG densities.  In such a case, the model can be extended by summing the current drive in each subband and incorporating intersubband scattering.  Effects such as double peak features in $g_{m}-V_{gs}$ plots are then expected to appear as higher subbands are populated.  


The authors acknowledge discussions with Patrick Fay, Huili Xing \& Tian Fang.  D.J. acknowledges financial support from DARPA NEXT (Dr. J. Albrecht) and S.R. from ONR (Dr. P. Maki).




\bibliographystyle{unsrt}


\pagebreak

\end{document}